# Mathematical and Physical Examination of the Locality Condition in Bell's Theorem

## H. Razmi

**Abstract**

*Using the Clauser–Horne model of Bell's theorem, the locality condition is examined, and it is shown that the corresponding formulation is equivalent to a factorization process consisting of three stages. The first stage is introduced based on the conditional probability definition in the classical theory of probability, and the other two stages are based on previously known relations called the outcome- and parameter-independence conditions.*

## 1. INTRODUCTION

It is a well-known fact that the Bell inequalities are not fully consistent with the predictions of quantum mechanics, and thus quantum theory should conflict with at least one of the assumptions used in the derivation of these inequalities (Bell's theorem). Since one of the main assumptions, among others (e.g., determinism and realism), used in the derivation of the Bell inequalities is the so-called locality condition and because there are some improvements to these inequalities, such as the Clauser–Horne (CH) inequality in which the assumption of determinism is eliminated,[1] the possible sources of the conflict in Bell's theorem have been narrowed to the locality condition. Furthermore, experimental predictions (e.g., the modern experiments of A. Aspect et al.[2,3]) also violate the Bell inequalities and are in good consistency with quantum physics. So the subject of quantum (non)locality has a great deal of importance in considering Bell's theorem.

Some important questions are introduced: "Is there a conflict between Einstein's locality principle and quantum theory by considering Bell's theorem? Does the experimental violation of the Bell inequalities imply nonlocality? How does quantum theory confront the special theory of relativity through Bell's theorem?" Although we have a powerful formulation of relativistic quantum field theory in the present standard model(s) of physics, the study of the above questions may reveal some important unknown facts about both the foundations of (quantum) physics and the real world around us.

An interesting publication[4] by A. Shimony studies the mathematical formulation of the locality assumption in detail and tries to show that the possible nonlocal properties seen in Bell's theorem don't mean there is a conflict between Einstein's locality principle and quantum theory. By the way, it seems we need to further study the mathematical formulation of the assumptions in Bell's theorem. Here, we want to improve the above-mentioned work[4] by studying the mathematical/physical formulation of the locality condition in more detail. In what follows it will be shown that the mathematical/physical formulation of the locality condition includes three stages of factorization: the first stage is based on the conditional probability definition in the classical theory of probability and the other stages are based on two previously known relations called the outcome- and parameter-independence conditions.

The model with which we work here is the CH model, which, in addition to being free of the assumption of determinism, is one of the closest models of Bell's theorem to the real world of experiments.

The standard Bell inequalities apply to a pair of spatially separated systems and are written in terms of correlations between measurable quantities associated with the two systems. Consider a system that decays into two spin-1/2 particles. The particles are produced in a singlet state (total spin = 0) and go in opposite directions. Each particle goes through a Stern–Gerlach apparatus and is then detected. The Stern–Gerlach apparatus receiving particle 1 takes orientation $\hat{a}$ or $\hat{a}'$, and the one receiving particle 2 takes orientation $\hat{b}$ or $\hat{b}'$.

Denote by $P_1(\hat{a}, \lambda)$ and $P_2(\hat{b}, \lambda)$ the probabilities of the detection of particles 1 and 2, respectively, and by $P_{12}(\hat{a}, \hat{b}, \lambda)$ the probability that both particles are detected simultaneously. Here $\lambda$ denotes the collection of (hidden) variables characterizing the state of each particle (the CH model is stochastic and realistic) with a normalized probability distribution

$$\int d\lambda \rho(\lambda) = 1. \tag{1}$$

In a seminal paper[1] Clauser and Horne derived the inequality

$$
\begin{aligned}
-1 \leq P_{12}(\hat{a}, \hat{b}) &- P_{12}(\hat{a}, \hat{b}') + P_{12}(\hat{a}', \hat{b}) \\
&+ P_{12}(\hat{a}', \hat{b}') - P_1(\hat{a}') - P_2(\hat{b}) \leq 0,
\end{aligned}
\tag{2}
$$

where $P_1(\hat{a})$, $P_2(\hat{b})$, and $P_{12}(\hat{a}, \hat{b})$ (similarly for primed angles) are the probabilities of detecting a count at the left detector (e.g., $D_1$), a count at the right detector (e.g., $D_2$), and a coincidence (simultaneous detection by both detectors), respectively. All these probabilities are averaged over the distribution function $\rho(\lambda)$ as

$$P_1(\hat{a}) = \int d\lambda \rho(\lambda) P_1(\hat{a}, \lambda), \tag{3}$$

$$P_2(\hat{b}) = \int d\lambda \rho(\lambda) P_2(\hat{b}, \lambda), \tag{4}$$

$$P_{12}(\hat{a}, \hat{b}) = \int d\lambda \rho(\lambda) P_{12}(\hat{a}, \hat{b}, \lambda). \tag{5}$$

The inequality (2) is the CH version of the Bell inequality. In deriving this inequality, Clauser and Horne used the locality condition

$$P_{12}(\hat{a}, \hat{b}, \lambda) = P_1(\hat{a}, \lambda) P_2(\hat{b}, \lambda) \tag{6}$$

to ensure that there is no action at a distance between instrument(s) 1 and instrument(s) 2.

Since we have assumed that the experiments involve dichotomic ($\pm$) parameters (e.g., spin-1/2 states), the probability functions $P_1(\hat{a}, \lambda)$, $P_2(\hat{b}, \lambda)$, and $P_{12}(\hat{a}, \hat{b}, \lambda)$ take the forms $P_1(\vec{\sigma}_1 \cdot \hat{a} = +1, \lambda)$ as the probability for the detection of particle 1 in its up (+) state, $P_2(\vec{\sigma}_2 \cdot \hat{b} = +1, \lambda)$ as the detection probability for particle 2 in its up (+) state, and $P_{12}(\vec{\sigma}_1 \cdot \hat{a} = +1, \vec{\sigma}_2 \cdot \hat{b} = +1, \lambda)$ as the probability for simultaneous detection of both particles in their up (+) states, respectively, so the locality condition (6) is rewritten as

$$
\begin{aligned}
P_{12}(\vec{\sigma}_1 \cdot \hat{a} &= +1, \vec{\sigma}_2 \cdot \hat{b} = +1, \lambda) \\
&= P_1(\vec{\sigma}_1 \cdot \hat{a} = +1, \lambda) P_2(\vec{\sigma}_2 \cdot \hat{b} = +1, \lambda).
\end{aligned}
\tag{7}
$$

We should mention that the introduction of the Pauli matrices ($\vec{\sigma}$) here doesn't mean we want to use quantum-mechanical formalism at this stage. What we mean by this formulation is that we are working with dichotomic ($\pm$) parameters. These parameters, regardless of the quantum-mechanical formalism, are experimentally well known (e.g., the proton-proton scattering experiment[5] or the experiments on correlations of linear polarizations of pairs of photons[2,3]). One can compare (identify) these parameters to the Stokes parameters used in the analysis of the polarization of electromagnetic waves and well known in classical electrodynamics and optics.

## 2. THE MATHEMATICAL/PHYSICAL FORMULATION OF THE LOCALITY CONDITION

As we know, there are different models (e.g., the original work of Bell[6]) of proving Bell's theorem with different forms of the locality assumption, so the mathematical formulation of the locality condition is (more or less) model dependent. Of course, they have one thing in common, that is, the factorization of a correlation function, as a product rule, into the multiplication of two functions (e.g., the factorization $P_{12} = P_1 \cdot P_2$ in the CH model). Careful attention to the explicit mathematical form of such product rules (the locality relations) leads us to find out that these mathematical relations implicitly consist of a number of stages of factorization. Based on our present knowledge we can show at least three stages needed for the factorization of $P_{12}$ in the locality relation (7) (i.e., three stages needed to get from the left-hand side of (7) to its right-hand side). As the first stage of this factorization, based on the definition of conditional probability in the classical theory of probability[7] (see the Appendix), we can write

$$
\begin{aligned}
P_{12}(\vec{\sigma}_1 \cdot \hat{a} &= +1, \vec{\sigma}_2 \cdot \hat{b} = +1, \lambda) \\
&= P_1(\vec{\sigma}_1 \cdot \hat{a} = +1, \lambda \,|\, a, b) \\
&\times P_2(\vec{\sigma}_2 \cdot \hat{b} = +1, \lambda \,|\, a, b, \vec{\sigma}_1 \cdot \hat{a} = +1),
\end{aligned}
\tag{8}
$$

where $P_1(\vec{\sigma}_1 \cdot \hat{a} = +1, \lambda \,|\, a, b)$ is the probability of the detection of particle 1 in its up (+) state when (if) the analyzers receiving particles 1 and 2 have orientations (parameter settings) $\hat{a}$ and $\hat{b}$, respectively, and $P_2(\vec{\sigma}_2 \cdot \hat{b} = +1, \lambda | a, b, \vec{\sigma}_1 \cdot \hat{a} = +1)$ is the detection probability for particle 2 in its up (+) state when (if) the analyzers are in the directions (having parameter settings) $\hat{a}$ and $\hat{b}$ and particle 1 is in its up (+) state. The notation is such that the vector characters (e.g., $\hat{a}$) correspond to the results (outcomes), while the scalar characters (e.g., $a$) correspond to setting angles (parameters).

The second stage of the factorization is a relation first introduced and named by Shimony[4] as the outcome-independence condition:

$$P_2(\vec{\sigma}_2 \cdot \hat{b} = +1, \lambda \,|\, a, b, \vec{\sigma}_1 \cdot \hat{a} = +1)$$
$$= P_2(\vec{\sigma}_2 \cdot \hat{b} = +1, \lambda \,|\, a, b), \qquad (9)$$

by means of which the relation (8) can be written as

$$P_{12}(\vec{\sigma}_1 \cdot \hat{a} = +1, \vec{\sigma}_2 \cdot \hat{b} = +1, \lambda)$$
$$= P_1(\vec{\sigma}_1 \cdot \hat{a} = +1, \lambda \,|\, a, b) P_2(\vec{\sigma}_2 \cdot \hat{b} = +1, \lambda \,|\, a, b), \qquad (10)$$

which is just the locality relation (7).

As the third stage of the factorization process, we use the following relations, which were introduced and named by Shimony[4] as the parameter-independence conditions:

$$P_1(\vec{\sigma}_1 \cdot \hat{a} = +1, \lambda \,|\, a, b) = P_1(\vec{\sigma}_1 \cdot \hat{a} = +1, \lambda) \text{ and}$$
$$P_2(\vec{\sigma}_2 \cdot \hat{b} = +1, \lambda \,|\, a, b) = P_2(\vec{\sigma}_2 \cdot \hat{b} = +1, \lambda), \qquad (11)$$

by means of which the relation (10) can be written as

$$P_{12}(\vec{\sigma}_1 \cdot \hat{a} = +1, \vec{\sigma}_2 \cdot \hat{b} = +1, \lambda)$$
$$= P_1(\vec{\sigma}_1 \cdot \hat{a} = +1, \lambda) P_2(\vec{\sigma}_2 \cdot \hat{b} = +1, \lambda), \qquad (12)$$

which is just the locality relation (7).

## 3. CONCLUSION

Since the concept of locality is important in a wide variety of subjects (e.g., in philosophy, it is related to the subject of causality; in the new subject of quantum information, it is related to quantum entanglement) and since one of the main assumptions in Bell's theorem is the locality condition, it is important to examine the mathematical/physical formulation of this concept to better understand both our physical theories (e.g., quantum theory) and the real world of phenomena around us.

Here we have considered the locality condition in the CH model of Bell's theorem to show that the corresponding mathematical/physical formulation is equivalent to a factorization process consisting of three stages. The first stage is based on the conditional probability definition in the classical theory of probability, and the other two stages are based on previously known relations called the outcome- and parameter-independence conditions.

Shimony[4] argued that one can assume that quantum mechanics is compatible with the parameter-independence condition(s) (11) and violates the outcome-independence condition(s) (9), and since only parameter independence is expected from the special theory of relativity, outcome independence may be considered to be the cause of the Bell inequality (here CH inequality) violation; thus there is a "peaceful coexistence" between quantum theory and the special theory of relativity.

One may assume that the source of conflict between quantum mechanics and the stochastic local realistic (CH) model is in the incompatibility of quantum theory with the first stage of the factorization process introduced here (i.e., the cause of the Bell (CH) inequality violation may be the incompatibility between quantum theory and the conditional probability definition in the classical theory of probability). Although this may be considered as an improvement to the above-noted "peaceful coexistence" between quantum theory and the special theory of relativity, it introduces new questions on the validity of the application of the classical theory of probability axioms/definitions to quantum theory. Indeed, other people have criticized the application of the axioms of the classical theory of probability in the derivation of the Bell inequalities. Kracklauer, citing Jaynes[8] as the first one to criticize the derivation of the Bell inequalities based on Bayes' formula (which is just the conditional probability definition introduced in this work), has discussed that Bell inequalities cannot be derived using Bayes' formula for conditional probabilities.[9,10] Hess and Philipp, also, have pointed out that the known proofs of Bell's inequalities contain algebraic manipulations that are not appropriate within the syntax of Kolmogorov's axioms for probability theory without detailed justification.[11] In our opinion, for the quantum-mechanical singlet state, the decomposition of the probability function $P_{12}$ into the form $P_1 \cdot P_2$ (regardless of the details in the functional forms of $P_1$ and $P_2$) is not possible; this is because the singlet state cannot be factored as a tensor product of its two parts/components (it is a nonfactorable state[12]). Such nonfactorability is directly related to the quantum entanglement property of the singlet state, a property that has an important role in the hot new subjects of quantum information and quantum computing.[13]

## APPENDIX

Consider a total sample space of parameters (all the parameters, including hidden variables and all instrumental setup variables consisting of angles) $\Omega_0$ that may be partitioned into two subspaces $\Omega_1$ and $\Omega_2$. $\Omega_1$ and $\Omega_2$ depend on the hidden variables and the setting parameters of particle 1 and/or 2. For example, $\Omega_1$ depends on $\lambda$, the setting parameters of particle 1, and maybe even the setting parameters of particle 2. The probability functions $P_1$, $P_2$, and $P_{12}$ are defined on the sample spaces $\Omega_1$, $\Omega_2$, and $(\Omega_1 \cap \Omega_2)$, respectively. If the information is that the sample lies in a known subset $\Omega_i$ ($i = 1, 2$), then the classical conditional probability is defined as

$$P(\Omega_i | \Omega_j) = \frac{P(\Omega_i \cap \Omega_j)}{P(\Omega_j)} \quad \text{(A1)}$$

$$\Rightarrow P(\Omega_i \cap \Omega_j) = P(\Omega_i | \Omega_j) P(\Omega_j),$$

where $j = (1, 2)$ is a free index similar to $i$, $P(\Omega_i | \Omega_j)$ is the (conditional ) probability of the event corresponding to the sample space $\Omega_i$ assuming (if) the event corresponding to the sample space $\Omega_j$ has occurred, and $P(\Omega_i \cap \Omega_j)$ is the correlation/joint probability of the event corresponding to the joint space $(\Omega_i \cap \Omega_j)$. As a particular example for the above definition and in order to justify the relation (8), assume $\Omega_1$ is the sample space of the event $\vec{\sigma}_1 \cdot \hat{a} = +1$ for an arbitrary hidden variable $\lambda$ when (if) the parameter setting of the left/right detector is in the $\hat{a}/\hat{b}$ direction and $\Omega_2$ is the sample space of the event of $\vec{\sigma}_2 \cdot \hat{b} = +1$ for an arbitrary hidden variable $\lambda$ when (if) the parameter setting of the left/right detector is in the $\hat{a}/\hat{b}$ direction. This means that the sample spaces $\Omega_1$ and $\Omega_2$ correspond to the events $(\vec{\sigma}_1 \cdot \hat{a} = +1, \lambda | a, b)$ and $(\vec{\sigma}_2 \cdot \hat{b} = +1, \lambda | a, b)$, respectively. Thus the probability function $P(\Omega_1)$, the conditional probability $P(\Omega_2 | \Omega_1)$, and the correlation/joint probability function $P(\Omega_1 \cap \Omega_2)$ in (A1) can be compared to the probability functions $P_1(\vec{\sigma}_1 \cdot \hat{a} = +1, \lambda | a, b)$, $P_2(\vec{\sigma}_2 \cdot \hat{b} = +1, \lambda | a, b, \vec{\sigma}_1 \cdot \hat{a} = +1)$, and $P_{12}(\vec{\sigma}_1 \cdot \hat{a} = +1, \vec{\sigma}_2 \cdot \hat{b} = +1, \lambda)$ in (8), respectively.

### Résumé

*En utilisant le modèle de Clauser–Horne du théorème de Bell, la condition de localité est examinée et on montre que la formulation correspondante est équivalente à un processus de factorisation se composant de trois étapes. La première étape est présentée basée sur la définition de la probabilité conditionnelle dans la théorie de probabilité classique et les deux autres étapes sont basées sur des relations précédentes connues appelées le résultat et les conditions d'indépendance des paramètres.*

## References

1. J.F. Clauser and M.A. Horne, Phys. Rev. D **10,** 526 (1974).
2. A. Aspect, J. Dalibard, and G. Roger, Phys. Rev. Lett. **47,** 460 (1981).
3. *Ibid.* **49,** 1804 (1982).
4. A. Shimony, *Sixty-two Years of Uncertainty*, edited by A.I. Miller (Plenum Press, New York, 1990), p. 33.
5. M. Lamehi-Rachti and W. Mittig, Phys. Rev. D **14,** 2543 (1976).
6. J.S. Bell, Physics **1,** 195 (1964).
7. R.T. Cox, *The Algebra of Probable Inference* (Johns Hopkins Press, Baltimore, MD, 1961).
8. E.T. Jaynes, *Maximum Entropy and Bayesian Methods*, edited by J. Skilling (Kluwer, Dordrecht, 1989), pp. 1–29.
9. A.F. Kracklauer, *Foundations of Probability and Physics*, 3, edited by A. Khrennikov, AIP Conference Proceedings **750,** 219 (AIP, Melville, NY, 2005).
10. *Idem*, arXiv/quant-ph/0602080.
11. K. Hess and W. Philipp, PNAS **101,** 1799 (2004).
12. F. Selleri, *Quantum Paradoxes and Physical Reality*, edited by Alwyn van der Merwe (Kluwer Academic, Dordrecht, Boston, 1990).
13. D. Bouwmeester, A.K. Ekert, and A. Zeilinger, *The Physics of Quantum Information* (Springer, Berlin, New York, 2000).

**H. Razmi**
Department of Physics
The University of Qom
Qom, I. R. Iran

e-mail: razmi@qom.ac.ir